\documentclass[twocolumn,showpacs,preprintnumbers,amsmath,amssymb]{revtex4}

\usepackage{dcolumn}% Align table columns on decimal point
\usepackage{bm}% bold math

\usepackage{latexsym}
\usepackage{epsfig}
\usepackage{amsmath}
\usepackage{amsfonts}
\usepackage{amsthm}
\usepackage{amssymb}
\usepackage{url}
\usepackage{hyperref}

\newcommand{\ket}[1]{\left | \, #1 \right \rangle}
\newcommand{\bra}[1]{\left \langle #1 \, \right |}
\newcommand{\proj}[2]{\ket{#1}\bra{#2}}

\newcommand{\braket}[2]{\left \langle #1 | #2 \right \rangle}

\newtheorem{theorem}{Theorem}

\begin{document}
\title{Simulating quantum systems using real Hilbert spaces}
\author{Matthew McKague}
	\affiliation{Institute for Quantum Computing and Dept. of Combinatorics \& Optimization \\ 
		University of Waterloo}
\author{Michele Mosca}
	\affiliation{Institute for Quantum Computing and Dept. of Combinatorics \& Optimization \\ 
		University of Waterloo and St. JeromeÕs University, \\ 
		and Perimeter Institute for Theoretical Physics}
\author{Nicolas Gisin}
	\affiliation{Group of Applied Physics, University of Geneva, CH-1211Geneva 4, Switzerland}

\preprint{preprintnum}

\date{\today}

\begin{abstract}
We develop a means of simulating the evolution and measurement of a multipartite quantum state under discrete or continuous evolution using another quantum system with states and operators lying in a real Hilbert space.  This extends previous results which were unable to simulate local evolution and measurements with local operators and was limited to discrete evolution.  We also detail applications to Bell inequalities and self-testing of quantum apparatus.

\end{abstract}

\maketitle

%%%%%%%%%%%%%%%%%%%%%%%%%%%%%%%%%%%%%%%%%%%%%%%%%%%%%%%%%%
\section{Introduction}
Coherence and interferometry are central to quantum physics, hence the fundamental importance of concepts like global and relative phases, single and multi-particle interferences, etc.. Phases are usually described by a complex number with norm one and the use of complex numbers seems intimately connected to the heart of quantum physics: the complex Hilbert space and the imaginary unit {\it i} in the Schr\"odinger equation. Still, we all know that the complex field $\mathbb{C}$ is isomorphic to a two dimensional real plane, and hence it is possible, at least formally, to compute all quantum predictions using only real numbers.  It is even well known that it is possible to simulate unitary evolution by a quantum system restricted to real amplitudes and matrix entries by doubling the dimension of the Hilbert space\footnote{This simulation has been in the folklore for a while, e.g. \cite{Rudolph:2002:A-2-rebit-gate-}}.  However, this simulation breaks down for multipartite systems.

Consider the case of two particles, first in a product state $\phi\otimes\psi$, and let's add a phase. This phase can equivalently be attributed to either of the two particles:
 \[
	\left( e^{i\alpha}\phi\right)\otimes \psi = \phi\otimes\left( e^{i\alpha} \psi \right).
\]
So far so good, $\alpha$ is only an (irrelevant) global phase, but this is no longer the case if the particles are in an entangled state. In such a case it is no longer clear to which of the two complex Hilbert spaces, the one describing the first particle or the second one, should be doubled in order to replace the imaginary part by additional real dimensions. It is even not obvious whether such a doubling could ``work locally''. By this we mean the following:  assume the particles, two or possibly more, are spatially separated from each other. In such a case one may ask whether one can double some or all the Hilbert spaces and let each party (each party holds one and only one of the particles) manipulate the phases independently of each other, including in the case of arbitrary entanglement.

Another situation in which the real simulation seems to break down is in the case of continuous time evolution.  It is generally understood that over time a state will evolve according to the Schr\"odinger equation, picking up a complex phase due to the $i$ in the exponent.  It would seem that, although discrete time evolution may work with only real numbers, continuous time evolution does not.

In this paper, we detail and expand the method in \cite{Matthew-McKague:2007:Simulating-Quan} which extends the well-known simulation using real Hilbert spaces to a more flexible simulation which requires only local access to ancilla qubits.  We also present a modification of the Schr\"odinger equation that allows one to view continuous time evolution as a process which happens over a real Hilbert space.  

We then present two particular applications where the multi-partite simulation answers open questions.  Magniez et al. \cite{Magniez:2006:Self-Testing-of} developed an algorithm for testing black box quantum devices allowing testing of quantum circuits, but the central theorem breaks down for unitaries with complex entries.  The current work shows that it is impossible to test 
arbitrary unitaries (in their black box model) since the real simulation is not unitarily equivalent to the original system, and yet it produces exactly the same measurement results.  Hence the theorem cannot be strengthened.  Also, we answer a question asked by Gisin \cite{Gisin:2007:Bell-inequaliti} about Bell inequalities.  He asked whether any Bell inequality could be maximally violated using states and measurement operators over real Hilbert spaces.  Our multi-partite simulation gives an affirmative answer to this question.

%%%%%%%%%%%%%%%%%%%%%%%%%%%%%%%%%%%%%%%%%%%%%%%%%%%%%%%%%%
\section{Simulating quantum systems}
\subsection{Simulating complex quantum states and operators}

Our goal is to simulate a given quantum system using only real numbers in state amplitudes and operator matrix entries.  This problem can be solved by adding one extra qubit and storing the two dimensions of the complex plane in the two dimensions of the extra qubit.  Specifically, we map states as follows (further discussion is given in the Appendix):

\begin{equation}\label{statemap}
	\sum_{x} (a_{x} + i b_{x}) \ket{x}
	\mapsto
	\sum_{x}a_{x}\ket{x}\ket{0} + b_{x}\ket{x}\ket{1}.
\end{equation}

\noindent In order for interference to happen properly, we need to combine the real and imaginary parts correctly.  Note that the matrix

\[
	XZ = \left(
	\begin{matrix}
		0 & -1 \\
		1 & 0 \\
	\end{matrix}
	\right)
\]

\noindent applied to the extra qubit corresponds to multiplying the original state by $i$.  With this in mind we see that the map

\begin{equation}\label{opmap}
	\sum_{x, x^{\prime}} (a_{x, x^{\prime}} + i b_{x,x^{\prime}}) \ket{x}\bra{x^{\prime}}
	\mapsto
	\sum_{x,x^{\prime}} \ket{x}\bra{x^{\prime}} \otimes (a_{x,x^{\prime}} I + b_{x,x^{\prime}}XZ)
\end{equation}

\noindent will give operators that map a simulation state of an input vector to the simulation state of the corresponding output vector.  Finally, for states $\ket{\psi}$ and $\ket{\phi}$ with corresponding simulation states $\ket{\psi^{\prime}}$ and $\ket{\phi^{\prime}}$ the inner products are related by

\begin{equation}\label{innerprod}
	\braket{\psi^{\prime}}{\phi^{\prime}} = \text{Re}\, \braket{\psi}{\phi}.
\end{equation}

\subsection{Measurements, Unitaries, and Hamiltonians}

Suppose that we have a positive semi-definite operator $P$ with corresponding simulation operator $P^{\prime}$ given as in equation~\ref{opmap}.  Equation~\ref{innerprod} reveals that $\bra{\psi}P\ket{\psi} = \bra{\psi^{\prime}}P^{\prime}\ket{\psi^{\prime}}$, hence $P^{\prime}$ is positive semi-definite.  This also establishes that POVM outcome probabilities will be the same for simulation states and POVM elements as for the original states and POVM elements.

For a unitary $U$, the corresponding operator $U^{\prime}$ will also be unitary, and any subsequent measurement outcomes using a simulated POVM will give also the same outcome probabilities as the original states, unitaries and POVM.  That is to say, we have

\[ \bra{\psi}U^{\dagger} P U \ket{\psi} = \bra{\psi^{\prime}}{U^{\prime}}^{\dagger} P^{\prime} U^{\prime} \ket{\psi^{\prime}} .\]

The same substitution will also work for Kraus operators of completely positive maps.

For an analogue of the Schr\"{o}dinger equation, we expand the context a little.  Let $H$ be a Hermition operator, then $iH$ has imaginary eigenvalues and $U = exp(iHt)$ will be unitary.  Let $H^{\prime}$ be the simulation operator corresponding to $H$.  Then $H^{\prime}$ is also Hermitian.  Consider $(I \otimes XZ)H^{\prime}$: like $iH$, it will have imaginary eigenvalues and hence $U^{\prime} = exp(I \otimes XZ)H^{\prime}t$ will be unitary.  By checking the Taylor expansion, we can see that $U^{\prime}$ has all real entries, so simulation states will continue to have all real amplitudes as they evolve continuously over time under this modified Schr\"{o}dinger equation.

Combined with the map to simulation states, the simulation operators properly evolve the simulation states to correspond with the evolution of the original system.  In particular, the evolved simulation states, combined with the simulation measurements, give the same statistics as the evolved original states and original measurements.

Interestingly, in this encoding more operations are possible than in the original complex description.  In particular, complex conjugation can be implemented by applying a $Z$ operator to the extra qubit.  This allows the simulation of arbitrary anti-unitaries as well as partial anti-unitaries which act only on a subspace, plus compositions of these operators with unitaries.  However partial transposition, i.e. anti-unitaries acting on only one factor of a tensor space, can't be simulated in this way.

\subsection{Local operations and multipartite systems}

An important aspect of the previous discussion is that the simulation of any operation involving an imaginary phase involves interacting with the one additional qubit.  This means that local operations will not in general remain local in the simulation.  We can overcome this difficulty by storing the additional qubit in a carefully constructed two-dimensional subspace stored in several qubits.  We then add one new qubit to each subsystem in the original system and use it to implement local operators in the simulation.

Define the following states on $k$ qubits with $y \in \{0,1\}^{k}$:

\[
	\ket{\overline{0}} = \sqrt{\frac{1}{2^{k-1}}}
	\sum_{h(y)\, \text{even}} (-1)^{\frac{h(y)}{2}  } \ket{y}
\]

\noindent and
\[
	\ket{\overline{1}} = \sqrt{\frac{1}{2^{k-1}}}
	\sum_{h(y)\, \text{odd}}  (-1)^{\frac{h(y)-1}{2}} \ket{y},
\]

\noindent (see the appendix for further discussion) where $h(y)$ is the number of ones in $y$.  Interestingly, applying $XZ$ on any one of the qubits takes $\ket{\overline{0}}$ to $\ket{\overline{1}}$ and $\ket{\overline{1}}$ to $-\ket{\overline{0}}$.  Thus a locally applied $XZ$ on a single qubit has the effect of applying a logical $XZ$ on the non-local subspace spanned by $\ket{\overline{0}}$ and $\ket{\overline{1}}$.  Now instead of using the logical $XZ$ in constructing operators, we can choose any convenient $XZ$ applied to a single qubit out of the $k$ that we have added.  Thus local unitaries, Hamiltonians and other operators, can be converted into local simulation operators.

%%%%%%%%%%%%%%%%%%%%%%%%%%%%%%%%%%%%%%%%%%%%%%%%%%%%%%%%%%
\section{Applications}

%%%%%%%%%%%%%%%%%%%%%%%%%%%%%%%%%%%%%%%%%%%%%%%%%%%%%%%%%%

\subsection{Self-Testing}
Magniez et. al. (see \cite{Magniez:2006:Self-Testing-of}) prove the following theorem.

\begin{theorem}[Self Testing\footnote{We have simplified the theorem for brevity.}]
Let $T$ be a unitary operator with real entries acting on a qubit.
Suppose that 
\begin{itemize}
	\item $\ket{\psi} \in A \otimes B$ is a physical state that simulates the maximally entangled pair of qubits $\ket{\Phi^+}$
	\item  $G_A \otimes I_B \ket{\psi}$ simulates $T \otimes I \ket{\Phi^+}$
	\item  $G_A \otimes G_B \ket{\psi}$ simulates $\ket{\Phi^+}$
\end{itemize}
then $G_{A}$ is equivalent to $T$.

\end{theorem}

By ``equivalence'' we mean that there is a unitary that maps the physical system to the logical system, plus some ancilla, and by ``simulates'' we means that the measurement outcome probabilities are identical with respect to some specified logical and given physical operators.

One can imagine that the Hilbert spaces $A$ and $B$ contain more information than just the logical qubit.  Is it possible for an adversary to use this additional information to build physical devices that simulate their logical description but are not equivalent to it?  If $T$, the logical description, has real entries then the theorem tells us that this is impossible, but we would like to know whether the theorem can be strengthened to include all unitaries.

Using the simulation developed above an adversary can construct the states $\ket{\overline{0}}$ and $\ket{\overline{1}}$ on two qubits and store them in extra dimensions of $A$ and $B$.  Whether applying $G_{A}$ on $A$ or $G_{B}$ on $B$ the adversary can perform the real simulation of $T$ using local operators that have the form required in the theorem.  Using this construction all the simulation states give the same measurement statistics as their ideal counterparts.  Note that our real simulation is not equivalent to the logical system since inner products are not preserved, as seen in equation \ref{innerprod}.  Thus an adversary can simulate a complex $T$ in a non-equivalent way and there is no stronger theorem (with the same assumptions) for complex $T$.  This is not a counter-example to the original theorem because, when the states have real amplitudes, the simulation is equal to the logical system in a product state with the ancilla systems.

\subsection{Bell Inequalities}

Gisin \cite{Gisin:2007:Bell-inequaliti} asked the question whether Bell inequalities could always be maximally violated by states and measurement operators on a real Hilbert space. Let us consider the case of two particles and limit ourselves to finite dimensional Hilbert spaces. In such a case the Schmidt decomposition guarantees that one can choose bases such that any state can be written using only real amplitudes: $\ket{\psi}=\sum_j r_j\ket{\phi_j}\ket{\psi_j}$. Consider now the most well known Bell inequality, due to CHSH.  Interestingly, for any such state, the optimal measurement to that violate the CHSH inequality are such that all eigenvectors of all observables appearing in the inequality can all be written in the Schmidt bases $\{\phi_j\}$ and $\{\psi_j\}$ using only real numbers \cite{Gisin:1992:Maximal-violati}. 

The above observations led some physicists to search for inequalities that either require complex numbers (in the state and/or eigenvectors of observables) to be violated, or for which the use of complex number would increase the maximal achievable violation (this would be nice: one could experimentally decide whether some observed correlations require complex Hilbert spaces or whether real ones suffice). The CHSH inequality requires only two measurement on each side. Hence the natural next step was to investigate inequalities with 3 measurements per site, but there is only one such inequality (up to symmetries) and again one can chose bases such that all amplitudes (state and eigenvectors) are real. Note that if one restricts to two dimensional Hilbert spaces this has a nice geometrical illustration: measurements with real-amplitude eigenvectors are represented on the Poincar\'e sphere as vectors all lying on a great circle. Recently, at last, a Bell inequality for which the optimal settings do not lie on a circle has been found \cite{Gisin:2007:Bell-inequaliti}. Moreover, it is elegant: on one side the 3 settings form a orthogonal triedre (like the xyz coordinates) and on the other side there are 4 settings on the edges of a tetrahedron. Hence, a particular case of the general question outline in the previous paragraph is: can this 3x4 setting Bell inequality be violated using only real amplitudes by the same amount as using complex numbers?

Navascu\'{e}s et al.\cite{Miguel-Navascues:2007:15}, using a simulation technique similar to the one presented in \cite{Matthew-McKague:2007:Simulating-Quan}, and P\'{a}l and V\'{e}rtesi in \cite{Pal:2008:Efficiency-of-h} proved that indeed real numbers are sufficient for maximal violation of all bipartite Bell inequalities.  The present work answers the question in the affirmative for inequalities involving any number of parties.  However, the required dimension increases by a factor of 2 with each additional system added.

%%%%%%%%%%%%%%%%%%%%%%%%%%%%%%%%%%%%%%%%%%%%%%%%%%%%%%%
\section{Conclusion}

We have shown that complex numbers are not required in order to describe quantum mechanical systems and their evolution, including continuous-time
evolution and the evolution of multipartite systems. Specifically, there is a simulation of such systems using only states and operators with real
coefficients and most importantly this can be done with the same assumptions about which subsystems are allowed to interact (that is, with the same local
structure).

We illustrated the non-trivial implications for self-testing quantum apparatus and for testing Bell inequalities.

\section{Acknowledgements}

We are grateful to Sandu Popescu for asking about the connection to Schr\"{o}dinger's equation and for helpful discussions on this topic.
\emph{This work was partially supported by NSERC, DTO-ARO, ORDCF, CFI, CIFAR, Ontario-MRI, CRC, OCE, QuantumWorks, MITACS, the Swiss NCCR-QP and European IP QAP.}

\bibliography{Global_Bibliography}

\appendix

\section{Where did these states come from?}

\subsection{Simulation states}
There is some ambiguity in \ref{statemap} due to the fact that we can multiply a state by a global phase without changing the behaviour of the system, but it does result in a different (and possibly orthogonal) simulation state.  We can resolve this difficulty by looking at the density operator.  A $n \times n$ density operator will be mapped to a $2n \times 2n$ operator with double the rank and trace 2.  To get a density operator we thus divide by 2.  For a  pure state density operator, the simulation density operator is rank 2 with eigenvalue 1/2 occurring with multiplicity 2.  Consider the following.  For a pure state $\ket{\psi} = \sum_{x}\left(a_x + i b_x\right) \ket{x}$ (with density operator $\rho = \proj{\psi}{\psi}$) the simulation density operator will be

 \[\rho^{\prime} = \frac{1}{2}\sum_{x,x^{\prime}} \proj{x}{x^{\prime}}\otimes \left( \left(a_{x}a_{x^{\prime}} +b_{x}b_{x^{\prime}}\right)I +  \left(b_{x}a_{x^{\prime}} -a_{x}b_{x^{\prime}}\right)XZ\right). \]

 \noindent This can be written as an equal ensemble of the states

  \[
 \ket{\phi_{1}} = \sum_{x} \ket{x} \otimes\left( a_{x}\ket{0} + b_{x}\ket{1} \right)
  \]

\[
 \ket{\phi_{2}} = \sum_{x} \ket{x}\otimes\left(- b_{x}\ket{0} + a_{x}\ket{1} \right).
  \]

  \noindent The first of these is the state given in equation~\ref{statemap}.  For a Hermitian operator $P$ a brief calculation shows that $\text{Tr}(P^{\prime}\rho^{\prime}) = \bra{\phi}P^{\prime}\ket{\phi}$ where $\ket{\phi}$ is any normalized linear combination of $\ket{\phi_{1}}$ and $\ket{\phi_{2}}$.  Thus we can replace $\rho^{\prime}$ with the pure state $\ket{\phi_{1}}$ and get the same outcome statistics for any POVM.

The two dimensional space of pure states that arises by applying the simulation map on a pure state corresponds to the possible global phases.  In fact, by multiplying the original state by a phase and applying equation~\ref{statemap} one arrives at a superposition of $\ket{\phi_{1}}$ and $\ket{\phi_{2}}$.  Thus arbitrarily choosing $\ket{\phi_{1}}$ instead of another state in the subspace is the same as arbitrarily fixing a global phase.

%%%%%%%%%%%%%%%%%%%%%%%%%
\subsection{Multipartite simulation states}
One way of deriving $\ket{\overline{0}}$ and $\ket{\overline{1}}$ is to use the stabilizer formalism.  Note that we want our two dimensional subspace to have the property
 \[-(XZ)_{j} (XZ)_{l}\ket{\psi} = \ket{\psi}
\]
for all $\ket{\psi}$ in the subspace.  This just says that applying a phase of $i$ independently to two different subsystems should be the same as applying a phase of $-1$.  There are $k-1$ independent operators of this form (for example, they can all be written as products of $-(XZ)_{1} (XZ)_{j}$ for various $j$) so there is a two dimensional subspace that is stabilized by these operators.  The states $\ket{\overline{0}}$ and $\ket{\overline{1}}$ form a convenient basis for this space.  Moreover, this is the unique subspace that has the desired property.

\end{document}